\documentclass[preprint,12pt,a4paper]{elsarticle}

\usepackage[T1]{fontenc}
\usepackage{lmodern}
\usepackage{amssymb}
\usepackage{hyperref}
\hypersetup{colorlinks=true,linkcolor=blue!60!black,citecolor=blue!60!black,urlcolor=blue!60!black}
\usepackage{listings}
\usepackage{xcolor}
\lstdefinestyle{compact}{ basicstyle=\ttfamily\footnotesize, breaklines=true,
keywordstyle=
\color{blue}
\bfseries, commentstyle=
\color{gray}
\itshape, stringstyle=
\color{orange}
, numbers=left, numberstyle=\tiny, frame=single, xleftmargin=1em, framexleftmargin=0.5em
}
\usepackage{tikz}
\usetikzlibrary{trees,arrows.meta,positioning}

\usepackage{tabularx}
\usepackage{booktabs}
\usepackage{makecell}
\usepackage{array}
\newcolumntype{L}[1]{>{\raggedright\arraybackslash}p{#1}}
\newcolumntype{C}[1]{>{\centering\arraybackslash}p{#1}}

\journal{SoftwareX}

\begin{document}
  \renewcommand{\labelenumii}{\arabic{enumi}.\arabic{enumii}}

  \begin{frontmatter}

    \title{GESIS Surf: An Open-Source %
    Infrastructure for Privacy-Preserving Longitudinal Web Browsing Data Collection}

    \author[label1]{Mario Alberto Ramirez\corref{cor1}}
    \ead{mario.ramirez@gesis.org}
    \cortext[cor1]{Corresponding author.}
    \author[label1]{Frank Mangold}
    \ead{frank.mangold@gesis.org}
    \author[label1,label2]{Sebastian Stier}
    \ead{sebastian.stier@gesis.org}
    \address[label1]{Department Computational Social Science,
    GESIS~-- Leibniz Institute for the Social Sciences,
    Unter Sachsenhausen 6--8, 50667 Cologne, Germany}
    \address[label2]{School of Social Sciences, University of Mannheim,
    A5, 6, 68159 Mannheim, Germany}

\begin{abstract}
Understanding how people navigate digital media environments requires behavioral data beyond survey self-reports. However, existing infrastructures for collecting such data reproducibly, sustainably, and with strong privacy guarantees remain limited: commercial solutions lack transparency and data governance, while academic tools are typically designed for narrow, short-term use cases and do not support general-purpose academic panel data collections.

We address this gap with GESIS Surf, a research infrastructure for large-scale longitudinal collection of web behavioral data in naturalistic environments. The system consists of a browser extension (Chrome and Firefox), a backend with a REST application programming interface (API), and a relational database. It passively captures user interactions, page metadata, and full HTML snapshots of each rendered page, embedding each observation within a hierarchical session model that preserves the structure of browsing behavior.

GESIS Surf implements privacy-by-design through strict opt-in participation, per-domain collection rules, and client-side data minimization, ensuring that sensitive data are controlled at the point of collection. The system has been deployed with more than 1,000 participants in a large-scale academic panel infrastructure that includes surveys of participants, and supports reproducible, content-level, and structurally rich analysis of digital behavior. By combining hierarchical session modeling, content-level capture, and integrated privacy enforcement within a unified system, GESIS Surf provides a general-purpose infrastructure for studying digital behavior in a reproducible manner over time.

    \end{abstract}

    \begin{keyword}
      web tracking \sep digital behavioral data \sep browser-based data collection \sep computational social science \sep research infrastructure \sep longitudinal data collection

    \end{keyword}
  \end{frontmatter}

  \section*{Metadata}

  \begin{table}[!h]
    \begin{tabular}{|l|p{5.8cm}|p{5.8cm}|}
      \hline
      \textbf{Nr.} & \textbf{Code metadata description}                               & \textbf{Metadata}                                                                                                                                                                                                                                                                                       \\
      \hline
      C1           & Current code version                                             & V2.0.0                                                                                                                                                                                                                                                                                                                    \\
      \hline
      C2           & Permanent link to code/repository used for this code version     & \url{https://github.com/gesiscss/gesis_surf_extension}            \\
      \hline
      C3           & Permanent link to Reproducible Capsule                           & Not applicable                                                    \\
      \hline
      C4           & Legal Code License                                               & MIT License                                                       \\
      \hline
      C5           & Code versioning system used                                      & git                                                               \\
      \hline
      C6           & Software code languages, tools, and services used                & TypeScript, Vite (browser extension); Python, Django REST Framework, PostgreSQL (backend); Elasticsearch, Logstash, Kibana (logging); Docker, Docker Compose (infrastructure) \\
      \hline
      C7           & Compilation requirements, operating environments \& dependencies & Extension: Node.js, Vite build toolchain. Backend: Docker and Docker Compose; PostgreSQL 14+; Elasticsearch 8+ (ELK stack for logging). \\
      \hline
      C8           & If available Link to developer documentation/manual              & \url{https://github.com/gesiscss/gesis_surf_extension/blob/main/README.md} \\
      \hline
      C9           & Support email for questions                                      & mario.ramirez@gesis.org                                           \\
      \hline
    \end{tabular}
    \caption{Code metadata}
    \label{codeMetadata}
  \end{table}

  \begin{table}[!h]
    \begin{tabular}{|l|p{5.8cm}|p{5.8cm}|}
      \hline
      \textbf{Nr.} & \textbf{(Executable) software metadata description}                                                                    & \textbf{Metadata}                                                                             \\
      \hline
      S1           & Current software version                                                                                               & V2.0.0                                                                                                          \\
      \hline
      S2           & Permanent link to executables of this version                                                                          & \url{https://github.com/gesiscss/gesis_surf_extension/releases/tag/v2.0.0}                                                 \\
      \hline
      S3           & Permanent link to Reproducible Capsule                                                                                 & Not applicable                                                                                                  \\
      \hline
      S4           & Legal Software License                                                                                                 & MIT License                                                                                                     \\
      \hline
      S5           & Computing platforms/Operating Systems                                                                                  & Google Chrome, Mozilla Firefox; Windows, macOS, Linux                                                           \\
      \hline
      S6           & Installation requirements \& dependencies                                                                              & Browser extension: Google Chrome $\geq$ 88 or Mozilla Firefox $\geq$ 109. Backend: Docker and Docker Compose.   \\
      \hline
      S7           & If available, link to user manual - if formally published include a reference to the publication in the reference list & \url{https://github.com/gesiscss/gesis_surf_extension/blob/main/README.md}                                      \\
      \hline
      S8           & Support email for questions                                                                                            & mario.ramirez@gesis.org                                                                                         \\
      \hline
    \end{tabular}
    \caption{Software metadata}
    \label{executableMetadata}
  \end{table}

  \section{Motivation and significance}

Internet use has become increasingly individualized and fragmented, spanning a diverse range of platforms, websites, and usage practices that vary across users and contexts. While surveys remain central to understanding attitudes, they are often inaccurate and systematically biased when it comes to measuring (digital) media use \cite{parry2021systematic, scharkow2016accuracy}. Understanding digital media use scientifically therefore requires precise measurements of digital behavior. To generate such insights, academics recruit participants and collect behavioral data with their informed consent through browser-based technologies, thereby enabling valid observations of digital activities over time \cite{stier2020integrating, wagner2025digitalbehavioral}. These approaches have been used to investigate news exposure, political communication, information-seeking behavior in health and environmental domains, and online polarization across computational social science \cite[e.g.,][]{mangold2025webtracking, Christner2022}.

Despite these advances, a central limitation remains the lack of a general-purpose infrastructure for collecting web-based behavioral data in a reproducible, scalable, and privacy-compliant manner. Existing approaches span multiple technical paradigms, including browser-based tracking tools operating within the user's browser (e.g., Eule~\cite{haim2019computational}, WebTrack~\cite{adam2025webtracking}), proxy-based systems routing traffic through intermediary servers (e.g., Roxy~\cite{menchen2012roxy}, Robin~\cite{bodo2017algorithmic, moller2019explaining}), and passive network-level measurement that infers behavior from traffic monitored at institutional networks or Internet service providers~\cite{holz2020cryptomining}. Commercial solutions are optimized for analytics and marketing purposes, offering limited transparency, restricted data access, and non-standardized outputs that hinder scientific reuse. Academic tools, in contrast, are typically developed for specific study designs or short-term deployments and do not generalize well to longitudinal, passive data collection in panel settings and broader, general-purpose research applications. These approaches also differ substantially in terms of data granularity, privacy architecture, and suitability for different topical research interests \cite{Christner2022, mangold2023infrastrukturen}.

As a result, three interrelated challenges arise. First, no general-purpose system supports passive data collection under conditions suitable for longitudinal panel-based research designs. Second, academic tracking tools vary in their privacy-preserving capabilities, while commercial solutions often provide limited transparency and control over data collection and governance. Third, existing approaches capture only partial behavioral signals: network-level measurements record cross-site activity \cite{holz2020cryptomining} but discard both the structural representation of browsing sessions and the content-level information required to reconstruct user trajectories and analyze exposure.

To situate GESIS Surf within this landscape, Table~\ref{tab:comparison} compares design features across the main categories of academic tracking approaches, based on the representative implementations discussed above. Each category optimizes for specific dimensions---such as traffic-level measurement or content-level capture---while their integration into longitudinal, panel-ready infrastructures remains limited.

\begin{table}[htpb]
\centering

\caption{Comparison of academic approaches to collecting web browsing data, based on representative implementations.}
\label{tab:comparison}
\setlength{\tabcolsep}{4pt}
\footnotesize
\renewcommand{\arraystretch}{1.2}

\begin{tabular}{L{3.2cm}C{2.0cm}C{2.4cm}C{2.4cm}C{2.4cm}}
\toprule
\textbf{Feature}
& \textbf{GESIS \newline Surf}
& \textbf{Proxy-based tools}
& \textbf{Browser-based tools}
& \textbf{Passive network measurement} \\
\midrule

Passive, naturalistic data collection & \checkmark & \checkmark & \checkmark & \checkmark \\
Longitudinal / panel-ready & \checkmark & $\times$ & $\times$ & $\times$ \\
Content-level capture (in situ) & \checkmark & $\times$ & $\sim$ & $\times$ \\
Hierarchical session modeling & \checkmark & $\times$ & $\times$ & $\times$ \\
Privacy-by-design (client-side control) & \checkmark & $\times$ & $\sim$ & $\times$ \\
General-purpose research use & \checkmark & $\times$ & $\times$ & $\times$ \\

\bottomrule
\end{tabular}

\parbox{\linewidth}{
\textit{Note}: $\sim$ indicates partial support or implementation-specific limitations (e.g., selective capture, post-hoc reconstruction, or lack of personalized, rendered content). Content-level capture refers to rendered HTML as presented to the user at the moment of exposure. Commercial tracking solutions are not included in this comparison, as they are typically designed for proprietary analytics contexts and do not provide the transparency, data access, or reproducibility required for academic research.
}

\end{table}

GESIS Surf addresses this gap with a panel-ready infrastructure for longitudinal web behavioral data collection in naturalistic environments. The system consists of a browser extension (Chrome and Firefox), a REST API backend, and a relational database with standardized schemas. It passively captures scroll events, click interactions, page metadata, and full DOM/HTML snapshots, embedding each observation within a five-level session hierarchy that preserves the relational structure of browsing behavior. Data collection is strictly opt-in, and privacy is enforced through per-domain rules and a temporary private mode that implements data minimization directly on the participant's device. A typical deployment involves provisioning participant accounts via the backend, installing the browser extension, and collecting behavioral data during ordinary browsing. Data are stored locally and synchronized securely to the backend, where they can be accessed via API endpoints or direct database queries.

GESIS Surf makes three contributions:
\begin{enumerate}
    \item \textbf{Hierarchical session modeling:} a five-level structure enabling relational reconstruction of complete browsing trajectories across windows, tabs, domains, and interaction events;
    \item \textbf{Privacy-by-design enforcement:} a policy architecture that applies attribute-level masking locally on the participant's device, ensuring that sensitive data are controlled before transmission;
    \item \textbf{Content-level capture:} an event pipeline that records rendered HTML in situ, enabling analysis of participants' actual content exposure and avoiding the biases inherent to retrospective web scraping of URLs.
\end{enumerate}

  \section{Software description}

  \subsection{Software architecture}

GESIS Surf is designed as a modular system consisting of a browser extension, a backend API, and a relational database. These components jointly implement three core capabilities: hierarchical session tracking, privacy-by-design enforcement, and content-level data capture. The system is further guided by four design imperatives: precise behavioral measurement, privacy preservation, long-term sustainability, and horizontal scalability, which shape all technical decisions from session management to domain-level data collection policies.

User interactions are captured by the browser extension, processed locally according to domain-specific privacy policies, and transmitted to the backend, where they are stored in a relational schema that reflects the session hierarchy.

The browser extension is implemented in TypeScript using Vite and uses a shared Manifest V3 architecture across Chrome and Firefox, with browser-specific adaptations where required. It comprises two functional layers. The \emph{background service worker} manages the global session lifecycle: it initializes session identifiers on start-up, coordinates child-session creation as windows and tabs open, and maintains the extension's responsiveness under Chrome's Manifest V3 service-worker lifecycle. Because Manifest V3 terminates idle service workers, GESIS Surf employs a heartbeat mechanism to maintain active collection during ongoing browsing sessions.

Session objects are managed in IndexedDB across their full lifetime: when a trackable unit opens (a window, tab, or domain visit), a record is created in IndexedDB with its composite UUID key, updated as events accumulate, and once the unit closes, the completed object is transmitted securely to the backend over HTTPS.

\begin{figure}[tbp]
    \centering
    \includegraphics[width=\textwidth]{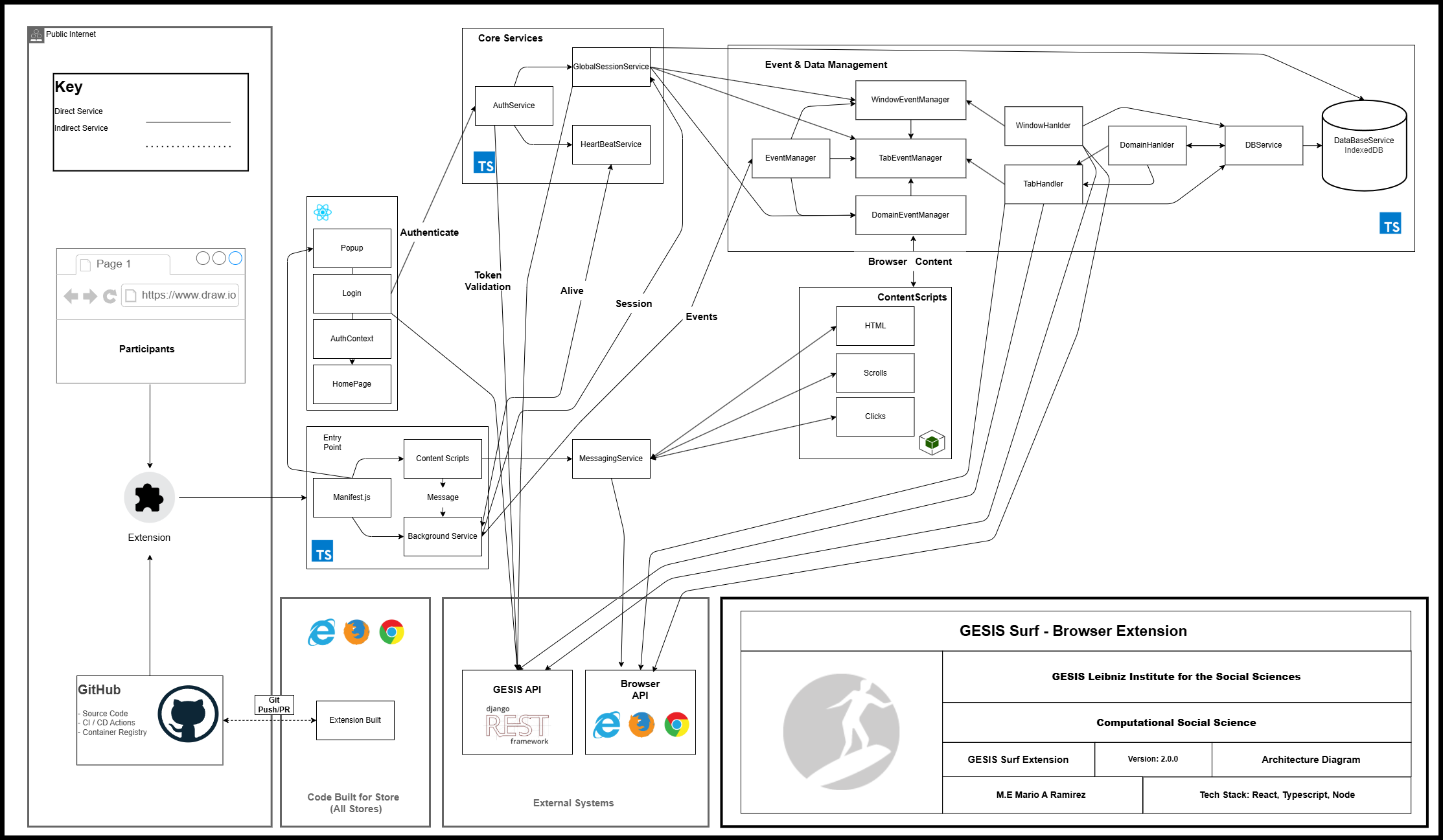}
    \caption{Client-side architecture of the GESIS Surf browser extension.
    The figure shows the internal extension components responsible for session
    management, data collection, local storage, and communication with the backend.}
    \label{fig:chrome_extension}
\end{figure}

The \emph{content script} is injected automatically into every visited page; it captures scroll events (position, page offsets, timestamp), click events (type, coordinates, target element, tag, identifier, class, referrer, timestamp), page metadata (title, description, favicon URL), and full DOM/HTML snapshots. Participants perform no additional task; capture is fully passive and imposes no visible interface on ordinary browsing. Figure~\ref{fig:chrome_extension} provides a client-side view of the GESIS Surf architecture, illustrating the internal components of the browser extension and the flow between its services.

  \subsubsection{Backend and infrastructure}

\begin{figure}[tbp]
    \centering
    \includegraphics[width=\textwidth]{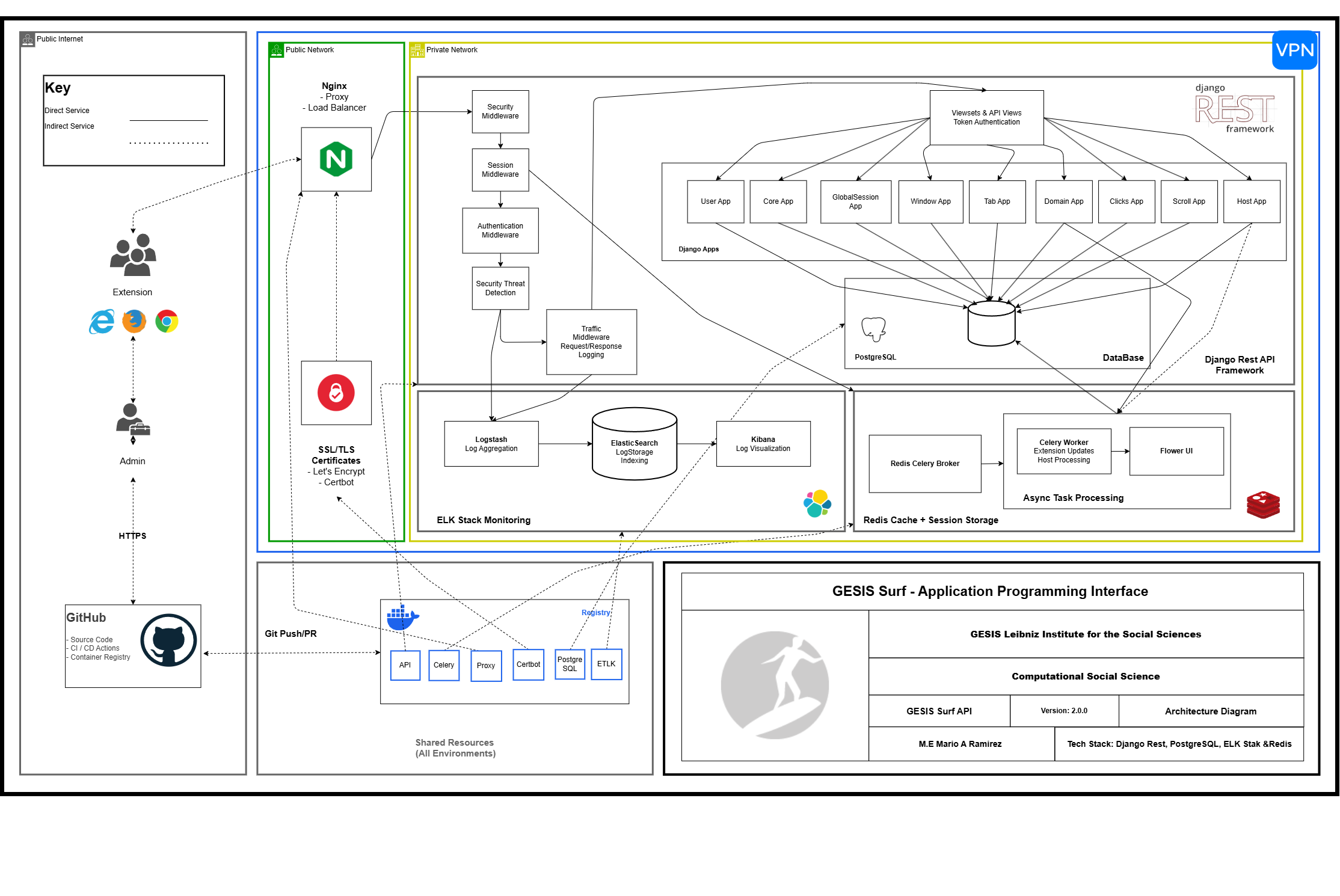}
    \caption{Backend and infrastructure architecture of GESIS Surf. The extension communicates
      over HTTPS through an Nginx reverse proxy into a private network containing
      the Django REST Framework application, PostgreSQL database, ELK stack
      monitoring, and Redis/Celery async task processing.}
    \label{fig:api_infra_stack}
  \end{figure}

The REST API is built with Django REST Framework and exposes token-authenticated endpoints organized around the five-level session hierarchy: dedicated Django apps handle global sessions, windows, tabs, domains, clicks, scrolls, and host rules. Whereas Figure~\ref{fig:chrome_extension} focuses on the client-side architecture of the browser extension, Figure~\ref{fig:api_infra_stack} provides a system-level view of the backend and supporting infrastructure.

Incoming payloads pass through a middleware chain (security, session, authentication, threat detection) before being routed to the corresponding data models. Data are stored in a normalized PostgreSQL relational schema whose entity relationships mirror the session containment model. This design ensures that the analytical representation of the data directly reflects the underlying behavioral structure captured at the client level. Researchers retrieve collected data either through authenticated REST API endpoints or by querying the PostgreSQL database directly.

GESIS Surf maintains two event pipelines. The \emph{domain pipeline} opens a new database record when the user navigates to a URL and closes the previous one, with the privacy classification applied before transmission. The \emph{content pipeline} forwards click, scroll, and HTML capture events---each evaluated and masked by the content policy service (Section~\ref{sec:privacy})---to dedicated REST endpoints; scroll events are throttled to one per second to limit bandwidth.

Asynchronous tasks---such as host-rule updates and extension processing---are handled by Celery workers backed by a Redis broker. System events feed into an ELK stack (Elasticsearch, Logstash, Kibana) that provides dashboards for operational monitoring, throughput analytics, and compliance documentation. Access logs are tamper-evident and stored in accordance with centrally defined institutional IT-security guidelines and technical and organizational measures. Every collected record also stores timestamps, the participant's active consent version, and the applied pipeline configuration, enabling full reproducibility of the collection context.

Docker ensures consistent environments across development, staging, and production. A GitHub Actions pipeline automatically executes tests, builds images, and deploys verified versions to an internal registry. Versioned releases guarantee traceability and rollback capability.

Together, these components provide a scalable and privacy-compliant infrastructure for longitudinal behavioral data collection, enabling content-level exposure analysis within a structured session framework.

  \subsection{Software functionalities}

GESIS Surf provides the following major functionalities:
\begin{itemize}
    \item passive capture of scroll and click events, page metadata, and HTML snapshots;
    \item hierarchical session tracking across browser windows, tabs, and domain visits;
    \item per-domain privacy policies and a temporary private mode, enforced on the participant's device;
    \item local buffering in IndexedDB with secure synchronization to the backend;
    \item data access via the REST API or direct SQL queries;
    \item operational monitoring and logging via an ELK stack.
\end{itemize}
The session model and the privacy policies are described in detail below.

  \subsubsection{Session model}

\begin{figure}[htbp]
    \centering
    \begin{tikzpicture}[
      every node/.style={
        draw, rounded corners=3pt, fill=gray!8,
        align=center, font=\small,
        minimum width=7.5cm, minimum height=0.65cm,
        inner sep=5pt
      },
      edge from parent/.style={draw, ->, >=stealth, thick},
      level distance=1.4cm,
      grow=down
    ]
    \node {\textbf{Global Session} \\ {\footnotesize one per browser-lifetime}}
      child { node {\textbf{Window Session} \\ {\footnotesize one per open browser window}}
        child { node {\textbf{Tab Session} \\ {\footnotesize one per browser tab}}
          child { node {\textbf{Domain Session} \\ {\footnotesize one per URL visited in a tab}}
            child { node {\textbf{Event} \\ {\footnotesize one per captured interaction}}
            }
          }
        }
      };
    \end{tikzpicture}
    \caption{Five-level session containment hierarchy. The hierarchy organizes browsing activity from global sessions to windows, tabs, domain visits, and individual interaction events.}
    \label{fig:session_hierarchy}
  \end{figure}
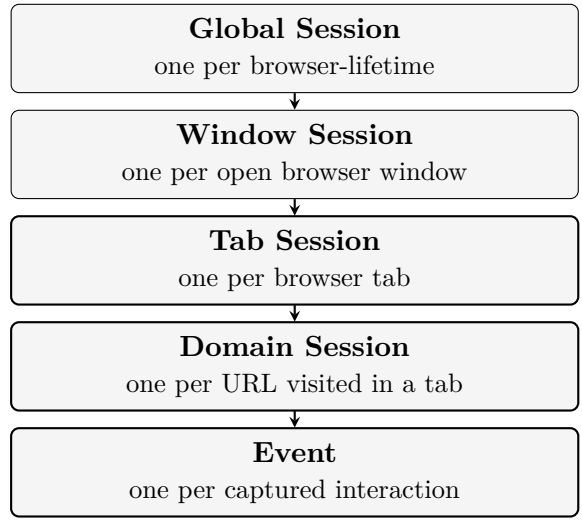

  GESIS Surf models each participant's web journey as a five-level hierarchy: \emph{global session} $\rightarrow$ \emph{browser window} $\rightarrow$ \emph{tab} $\rightarrow$ \emph{domain visit} $\rightarrow$ \emph{interaction event} (Figure~\ref{fig:session_hierarchy}). Each level is identified by a deterministic composite key that encodes its full ancestry, enabling researchers to reconstruct any segment of the web journey and query behavioral data at any level of granularity without ambiguity. This hierarchical structure enables the reconstruction of complete browsing trajectories and supports analysis across multiple levels of aggregation.

  Containment is strict: if any parent session is missing from IndexedDB when a child event is to be created, the event is discarded and the error is logged. Table~\ref{tab:session_ids} details the composition formula for each level. By preserving the relational structure of browsing behavior across windows, tabs, and domains, the model allows researchers to move beyond flat event logs or domain-level aggregates.

\begin{table}[htbp]
\caption{Session ID composition rules across the five-level hierarchy.}
\label{tab:session_ids}

\centering
\small
\begin{tabular}{ll}
\toprule
\textbf{Session level} & \textbf{Composition formula} \\
\midrule

Global Session
  & \texttt{global-session-\{UUID\}} \\

Window Session
  & \texttt{\{globalSessionId\}-windowId-\{windowId\}} \\

Tab Session
  & \texttt{\{windowSessionId\}-tabId-\{tabId\}} \\

Domain Session
  & \texttt{\{tabSessionId\}-domain-\{urlPart\}} \\

Event
  & References \texttt{domain\_id} (server-assigned UUID) \\

\bottomrule
\end{tabular}

\vspace{0.5em}
\parbox{\linewidth}{
\textit{Note}: The \texttt{\{urlPart\}} component of the Domain Session key is
policy-sensitive: full URL under \emph{full allow}/\emph{default},
hostname only under \emph{only host}, literal \texttt{"full\_deny"}
under \emph{full deny}.
}
\end{table}

\subsubsection{Privacy and data collection policies}\label{sec:privacy}

A server-managed host classification system governs collection on a per-domain basis. Each domain is assigned one of five privacy modes: \emph{default} (no explicit rule exists for the URL; full behavioral capture applies as fallback); \emph{full allow} (explicit full-collection rule: scroll events, click events, DOM snapshot, and metadata); \emph{only host} (URL restricted to the hostname, with sensitive content attributes masked and HTML capture disabled while retaining non-content interaction signals; e.g., \texttt{mail.google.com}); \emph{full deny} (domain and content information suppressed while retaining non-content interaction signals; e.g., \texttt{pornhub.com}); or \emph{private} (user-activated temporary suspension with 15-minute auto-expiry).

This design enforces data minimization directly at the point of data collection, rather than as a post-hoc filtering step, allowing sensitive browsing contexts to be excluded by design. Participation is strictly opt-in and the extension becomes active only after informed consent is recorded and after a user logs in with an individual user name and password. Table~\ref{tab:privacy_classifications} details the domain-level masking applied under each mode; Table~\ref{tab:collection_matrix} shows the resulting collection matrix at the content-event level.

  \begin{table}[htbp]
    \centering
    \small
    \caption{Privacy policy classifications and domain-level attribute masking.}
    \label{tab:privacy_classifications}

    \resizebox{\linewidth}{!}{%
    \begin{tabular}{lllll}
      \toprule
      \textbf{Classification} & \textbf{When applied}
        & \textbf{\texttt{domain\_url} sent to API}
        & \textbf{\texttt{domain\_title}}
        & \textbf{\texttt{domain\_fav\_icon}} \\
      \midrule
      \texttt{default} / \texttt{full\_allow}
        & Fallback / explicit full-collection rule
        & Full URL (path + query) & Real title & Real icon \\
      \texttt{only\_host}
        & Blocklist entry---hostname only
        & Hostname (e.g., \texttt{bank.example.com})
        & Masked (\texttt{"only\_host"}) & Masked \\
      \texttt{full\_deny}
        & Blocklist entry---fully suppressed
        & \texttt{"full\_deny"} literal
        & \texttt{"full\_deny"} & \texttt{"full\_deny"} \\
      \texttt{private}
        & Private Mode active
        & \texttt{"Private-Mode"} literal
        & \texttt{"Private-Mode"} & \texttt{"Private-Mode"} \\
      \bottomrule
    \end{tabular}}%
\end{table}

\begin{table}[htbp]
    \centering
    \small
    \caption{Content event data collection by privacy policy classification.}
    \label{tab:collection_matrix}

    \resizebox{\linewidth}{!}{%
    \begin{tabular}{lllll}
      \toprule
      \textbf{Classification}
        & \textbf{Click \texttt{referrer} + \texttt{target\_element}}
        & \textbf{Click coordinates (\texttt{x}, \texttt{y})}
        & \textbf{Scroll metrics}
        & \textbf{HTML snapshot} \\
      \midrule
      \texttt{default} / \texttt{full\_allow}
        & Original & Original & Original & Original \\
      \texttt{only\_host}
        & Masked $\rightarrow$ \texttt{"only\_host"} & Original & Original & Blocked \\
      \texttt{full\_deny}
        & Masked $\rightarrow$ \texttt{"full\_deny"} & Original & Original & Blocked \\
      \texttt{private}
        & Blocked & Blocked & Blocked & Blocked \\
      \bottomrule
    \end{tabular}}%
\end{table}

 Policy evaluation is implemented as a three-service hierarchy. A \texttt{Base\allowbreak Policy\allowbreak Service} resolves the active classification for any URL---checking Private Mode first, consulting cached host rules, then falling back to \texttt{default}---and two specialized services inherit from it: \texttt{DomainPolicyService}, which builds the domain payload with masking applied per Table~\ref{tab:privacy_classifications}, and \texttt{Content\allowbreak Policy\allowbreak Service}, which evaluates each event kind (click, scroll, HTML snapshot) and returns allow, mask, or block.

 All policy evaluation occurs locally on the user’s device before any network transmission, ensuring that sensitive data do not leave the client environment regardless of backend configuration. Host rules are fetched at login and cached locally in both IndexedDB and \texttt{chrome.storage.local}; updates are synchronized only when the rule version changes, reducing unnecessary network overhead.

  \section{Illustrative examples}

Building on the system architecture described above, the following example demonstrates a typical data collection and retrieval workflow using GESIS Surf, showing how hierarchical session tracking and privacy policies are applied during runtime. The example traces the full process from session initialization to data storage and retrieval, highlighting how behavioral data are structured, filtered, and made accessible for analysis:

  \begin{enumerate}
    \item \textbf{Initialization}. When the browser is opened, the extension initializes a global session via the \texttt{GlobalSessionService}. A corresponding window session is created, establishing the first levels of the session hierarchy before any navigation occurs.
    \item \textbf{Data collection under domain policies.} When the user navigates to a website (e.g., \texttt{domain\_url = "example-bank.com"}), the system evaluates the domain against the configured policy rules. In the case of an \texttt{"only\_host"} classification, the domain URL is reduced to the hostname, while metadata fields such as \texttt{title} and \texttt{favicon} are masked. Selected interaction-event attributes (e.g., click referrer and target element) are masked according to the policy, and content-level data such as HTML snapshots are not collected. All policy decisions are applied locally on the user's device prior to transmission.
\item \textbf{Session transitions and full capture.} Upon navigation to another domain (e.g., \texttt{"example-news.org"}), the system closes the previous domain session and creates a new one. If no restrictive policy applies, the default mode enables full data capture, including complete URLs, interaction events, and HTML snapshots. All events are stored within the hierarchical session structure, preserving their relational context across windows, tabs, and domains.

    \item \textbf{Data retrieval.}
    Collected data can be accessed either via the REST API or directly from the PostgreSQL database. For example, click events can be retrieved and joined with domain-level information using a structured query (Listing~\ref{lst:clickquery}).

\begin{lstlisting}[language=SQL,
  caption={Retrieving a participant's click events joined with their domain-level context.},
  label={lst:clickquery}]
SELECT c.click_time, c.click_page_x, c.click_page_y,
       c.click_target_tag, c.click_referrer,
       d.domain_url, d.start_time
FROM   clicks_click   c
JOIN   tab_domain     d ON d.id = c.domain_id
WHERE  c.user_id = '<participant_uuid>'
ORDER  BY c.click_time;
\end{lstlisting}

This query returns both masked and unmasked events depending on the applied domain policies, while preserving their position within the session hierarchy.

  \end{enumerate}

  \section{Impact}

GESIS Surf extends the capabilities of digital behavior research by enabling content-level and structurally rich behavioral data collection in naturalistic settings. By capturing the rendered HTML at the moment of exposure, the system allows researchers to move beyond domain-level proxies---recording only which websites were visited---towards analyzing the actual content encountered by users in situ. This addresses a central limitation of retrospective URL-based scraping, which distorts content measures mainly because content retrieved outside the participant's browser differs from what they saw (e.g., paywalls, login states, personalized elements), and only secondarily because of time delays \cite{ulloa2025scraping}.

The software further enables fine-grained reconstruction of browsing trajectories through its five-level session hierarchy. By explicitly modeling the nested relationships between windows, tabs, domain visits, and interaction events, GESIS Surf supports the analysis of cross-domain navigation, platform switching, and multi-tab browsing behavior within coherent user sessions. This structured representation allows queries at multiple levels of granularity while preserving contextual relationships, thereby extending the analytical scope of behavioral data beyond flat event logs or domain-level aggregates.

A key capability of the system is its integration with longitudinal survey panel infrastructures, allowing behavioral data to be linked with self-reported measures such as demographics and attitudes. This enables combined analyses of observed behavior and personal characteristics within the same individuals, supporting a more comprehensive and behaviorally grounded understanding of digital media use \cite{stier2020integrating}.

Beyond data collection, the system operationalizes privacy-by-design as a core architectural principle. Data minimization is enforced locally on the participant's device, ensuring that sensitive data are controlled prior to transmission. In contrast to tracking infrastructures that rely on server-side data collection or opaque data governance, GESIS Surf embeds privacy and ethics safeguards directly into the technical design, in line with calls to build robust ethics tools into user-sourced data collection \cite{meyer2023ethics} and to treat privacy implementation as a core criterion for tracking tools \cite{Christner2022}. This approach is supported by collaboration with data-protection officers and formal documentation (e.g., records of processing activities under the EU General Data Protection Regulation, GDPR), integrating compliance requirements as reusable components of the system architecture.

GESIS Surf is designed as a production-grade infrastructure rather than a project-specific prototype. The system has been deployed at scale within the GESIS Panel.dbd, involving more than 1,000 participants over a period of more than 2 years \cite{gilsbach2025incentives, gesis2025panel}. This deployment demonstrates the feasibility of long-term passive data collection for capturing user behavior in academic research. The implementation builds on experience from prior user-centric tracking studies with browser plug-ins and browsing panels \cite{breuer2023facebook, stier2022postbroadcast, haim2019computational} and addresses current technical constraints through local buffering via IndexedDB, compatibility with Chrome Manifest V3, and robust session management.

Finally, GESIS Surf supports reproducibility and long-term reuse through open-source availability (MIT license), semantic versioning, automated testing pipelines, and containerized deployment, allowing other researchers to independently deploy and adapt the software to their own research settings. Continuous maintenance by professional research software engineers, which is considered essential for research software sustainability \cite{anzt2021rse}, keeps the extension compatible with evolving browser environments, while the integration of governance, monitoring, and version control mechanisms enables transparent and replicable data collection processes. Together, these features position GESIS Surf as a scalable and reusable infrastructure for computational social science research on digital behavior.

  \section{Conclusions}

GESIS Surf provides an institutionally governed system for longitudinal digital behavioral data collection, combining fine-grained measurement with integrated data governance. The system thereby enables a shift from domain-level tracking to content-level and structurally rich behavioral analysis within a GDPR-compliant framework. It has been successfully deployed at scale within the GESIS Panel.dbd. As an open-source, continuously maintained infrastructure with integrated governance and reproducibility features, GESIS Surf enables the systematic study of content exposure and web browsing trajectories in naturalistic environments at scale.

  \section*{Acknowledgements}
  The authors thank Judith Gilsbach and Felix Schmidt for their contributions to the
  project, and Fernando Guzm\'an for his work as software architect on the first version
  of GESIS Surf.

  \section*{CRediT author contribution statement}
  \begin{description}
    \item[\textbf{Mario Alberto Ramirez:}] Conceptualization, Software, Methodology,
      Writing -- original draft.
    \item[\textbf{Frank Mangold:}] Writing -- review \& editing.
    \item[\textbf{Sebastian Stier:}] Writing -- review \& editing.
  \end{description}

  \section*{Funding}
  This research did not receive any specific grant from funding agencies in the public, commercial, or not-for-profit sectors.

  \section*{Declaration of competing interest}
  The authors declare that they have no known competing financial interests or personal relationships that could have appeared to influence the work reported in this paper.

\section*{Data availability}
No data were used for the research described in this article. Previously collected web browsing data obtained using GESIS Surf are available through the GESIS Data Archive at \url{https://doi.org/10.4232/1.14467}~\cite{gesis2025panel} and \url{https://doi.org/10.4232/1.14646}~\cite{gesis2025campus}.

  \section*{Declaration of generative AI and AI-assisted technologies in the manuscript preparation process}
  During the preparation of this work the authors used Claude (Anthropic) and ChatGPT (OpenAI) to improve clarity and language of the manuscript. After using these tools, the authors reviewed and edited the content as needed and take full responsibility for the content of the published article.

  \bibliographystyle{elsarticle-num}
  \bibliography{main}
\end{document}